# A Comparative Genomic Analysis of Coronavirus Families Using Chaos Game Representation and Fisher-Shannon Complexity

S. K. Laha

CSIR-Central Mechanical Engineering Research Institute

Durgapur, West Bengal, India, PIN-713209

Email: sk_laha@cmeri.res.in

Mobile No. 9749172042

**Abstract**

From its first emergence in Wuhan, China in December, 2019 the COVID-19 pandemic has caused unprecedented health crisis throughout the world. The novel coronavirus disease is caused by severe acute respiratory syndrome coronavirus 2 (SARS-CoV-2) which belongs to the coronaviridae family. In this paper, a comparative genomic analysis of eight coronaviruses namely Human coronavirus OC43 (HCoV-OC43), Human coronavirus HKU1 (HCoV-HKU1), Human coronavirus 229E (HCoV-229E), Human coronavirus NL63 (HCoV-NL63), Severe acute respiratory syndrome coronavirus (SARS-CoV), Middle East respiratory syndrome-related coronavirus (MERS-CoV), Severe acute respiratory syndrome coronavirus 2 (SARS-CoV-2) and Bat coronavirus RaTG13 has been carried out using Chaos Game Representation and Fisher-Shannon Complexity (CGR-FSC) measure. Chaos Game Representation (CGR) is a unique alignment-free method to visualize one dimensional DNA sequence in a two-dimensional fractal-like pattern. The two-dimensional CGR pattern is then quantified by Fisher-Shannon Complexity (FSC) measure. The CGR-FSC can effectively identify the viruses uniquely and their similarity/dissimilarity can be revealed in the Fisher-Shannon Information Plane (FSIP).

# 1. Introduction

Severe acute respiratory syndrome coronavirus 2 (SARS-CoV-2) virus causes severe respiratory illness along with various symptoms like fever, cough, headache, loss of smell etc. [1-2]. The novel coronavirus disease or COVID-19 disease has caused an unparallel global health crisis which has been declared as a pandemic by World Health Organization (WHO). SARS-CoV-2 is a new member of the coronaviridae family. Coronaviruses have distinct corona-like spike proteins emerging from the surfaces which enter their host cells by binding to the Angiotensin-converting enzyme 2 (ACE2) receptor [3]. The coronaviridae family is single-stranded, positive sense, RNA (+ssRNA) virus with an approximate genome size of 27-32 kb [4-5]. The family can be grouped into four genera: alpha, beta, gamma and delta [3].

Genomic signal processing (GSP), a subfield of Bioinformatics is a cross-disciplinary research topic at the intersection of genomics, computer science, statistics etc. Many biomolecular sequences such as DNA/RNA, proteins etc. can be converted into numeric strings before further downstream analysis. DNA/RNA is one such biomolecular sequence that consists of four nucleotides represented by four letters A, G, C and T/U which stands for Adenine, Guanine, Cytosine and Thymine/Uracil respectively. There are various encoding schemes developed for genomic data analysis such as atomic representation [7], chaos game representation [8], thermodynamic properties [9], Voss representation [10], DNA Walk model [11], H-Curve and Z-curve [12-13], 4D-Dynamic Representation of DNA/RNA Sequences [14] etc. The representation can be both graphical and non-graphical [15]. For a comprehensive review of such encoding schemes, authors can refer to [16].

Chaos Game Representation (CGR) proposed by Jeffrey [8] is one of the most popular encoding schemes. The 2D CGR of DNA sequence provides alignment-free fractal structure in a unit square matrix. CGR has been applied in the analysis of DNA sequence [17, 20-22] as well as protein structure [18-19]. Hoang et al. [17] studied the phylogenetic relationship using 2D CGR and the power spectrum of the corresponding DNA sequence. They obtained the phylogenetic tree of human rhinovirus, influenza virus and HPV using the above method and UPGMA. Sengupta et al. [20] carried out similarity studies of coronaviruses based on k-th order frequency chaos game representation (FCGR). Barbosa et al. [21] obtained the CGR maps of SARS-Cov-2 along with Betacoronavirus RaTG13, bat-SL-CoVZC45, and bat-SL-CoVZXC21. Deng and Luan [22] also used CGR in combination with Hurst exponent to study the similarity/dissimilarity of DNA sequences. They used their method on the first exon of $\beta$-globin gene of different species. Once the CGR maps are obtained, the next step is then to characterize the complexity pattern of the 2D signal obtained from the CGR process. Researchers have developed various methods to measure the time series signal complexity such as fractal dimension [23], entropies [24], Lyapunov exponent [25] etc. Fisher Information Measure (FIM), originally introduced by Fisher [26] for statistical estimation problems has been shown by Frieden [27] that FIM can be used to measure the degree of disorder of a physical system. Martin et al. [28] have shown that FIM can effectively capture the change in the dynamic behavior of nonlinear systems such as the logistic map, the Lorenz system and the Rossler system. Dembo et al. [29] in a classical paper studied the information-theoretic inequalities including Shannon Entropy Power (SEP) and Fisher Information Measure (FIM). Vignat and Bercher [30] introduced the concept of Fisher-Shannon Information Plane (FSIP) in which they have shown that simultaneous examination of both FIM and SEP can characterize a complex, nonlinear signal. The product of FIM and SEP called Fisher-Shannon Complexity (FSC) has been used to study high-frequency wind speed data in geosciences [31], the evolution of the daily maximum surface temperature distributions [32] and the time series data of Standardized Precipitation Index (SPI) [33].

In this paper, the Fisher-Shannon information approach has been applied for a comparative genomic analysis of eight coronaviruses namely Human coronavirus OC43 (HCoV-OC43), Human coronavirus HKU1 (HCoV-HKU1), Human coronavirus 229E (HCoV-229E), Human coronavirus NL63 (HCoV-NL63), Severe acute respiratory syndrome coronavirus (SARS-CoV), Middle East respiratory syndrome-related coronavirus (MERS-CoV), Severe acute respiratory syndrome coronavirus 2 (SARS-CoV-2) and Bat coronavirus RaTG13 (Bat-CoV). Out of these viruses Bat coronavirus, RaTG13 causes infection in horseshoe bat, *Rhinolophus affinis,* whereas the remaining seven viruses are known to cause infection in humans.

**2.1 Chaos Game Representation**

Chaos Game Representation (CGR) is an interesting process through which one-dimensional sequences can be converted to a two-dimensional space by an iterated function system (IFS). CGR, an alignment-free method, was proposed by Jeffrey [8] to visualize DNA sequences. The resulting two-dimensional representation is in the form of a scatter plot and many remarkable fractal-like patterns can be observed which are difficult to observe in the initial 1D sequences such as DNA, RNA, or protein sequences.

The 2D planar CGR space is a continuous unit square described by four vertices assigned by the four nucleotides i.e., Adenine (A), Guanine (G), Cytosine (C) and Thymine (T). In other words, the coordinates of these four nucleotides are given by A = (0, 0); T = (1, 0); G = (1, 1) and C = (0, 1). In this Cartesian plane any nucleotide sequence of any length can be uniquely determined. The CGR coordinates are determined iteratively by the following process: the first nucleotide position is halfway between the starting point and the corresponding vertex of the nucleotide where the starting point is at (0.5, 0.5). The successive nucleotides are then plotted halfway between the previous nucleotide position and the vertex representing that nucleotide. For a DNA sequence, the equation for the above iterative function system (IFS) is given by

$$X_i = 0.5(X_{i-1} + g_{ix})$$
$$Y_i = 0.5(Y_{i-1} + g_{iy})$$
(1)

where $(g_{ix}, g_{iy})$ is the corresponding vertex the current nucleotide whereas $(X_{i-1}, Y_{i-1})$ and $(X_i, Y_i)$ are the previous and current coordinates respectively. The iteration proceeds till the last nucleotide in the DNA sequence. Thus, from the CGR two series along the X and Y coordinates are obtained which are denoted by CGR-X and CGR-Y respectively.

As suggested by Deng and Luan [22], a one dimensional CGR-walk sequence can be obtained as a summation of the X and Y coordinates of the CGR map, i.e.

$$\text{CGR-Walk} = \text{CGR-}X + \text{CGR-}Y \quad (2)$$

where CGR-X and CGR-Y are the X and Y coordinates of the CGR map respectively.

## 2.2 Fisher-Shannon Analysis

Let X be a continuous univariate random variable with a probability density function denoted by $p_X(x)$. The differential entropy of $x$ is given by,

$$H_X = -\int p_X(x) \log p_X(x) dx \quad (3)$$

However, as suggested by Dembo et al. [29], it is sometimes convenient to express the above quantity in terms of Shannon Entropy Power (SEP), which is given by,

$$N_X = \frac{1}{2\pi e} e^{2H_X} \quad (4)$$

Further, the quantity Fisher Information Measure (FIM) is expressed as,

$$I_X = \int \frac{\left[\frac{\partial p_X(x)}{\partial x}\right]^2}{p_X(x)} \, dx \qquad (5)$$

It should be noted that FIM is conceptually different from Fisher information which is the information content of the random variable $X$ about its distribution parameters, $\theta$.

Both the FIM and SEP have been applied to study signal complexity. The Fisher-Shannon Complexity (FSC) can be defined as the product of FIM and SEP, as given by,

$$C_X = N_X I_X \qquad (6).$$

It can be shown that, $C_X \geq 1$, where the equality holds if and only if $X$ has Gaussian distribution [29].

## 3. Methods and Data

The whole-genome reference sequences of the viruses in the present study are downloaded from the NCBI Genbank. The downloaded genome sequences of the viruses are in the Fasta format and their Accession IDs are given in the following Table.

**Table 1**: Virus details

| Sl. No. | Virus | Accession ID | Type | Remark |
|---|---|---|---|---|
| 1 | Human coronavirus OC43 (HCoV-OC43) | NC_006213.1 | β-CoV | |
| 2 | Human coronavirus HKU1 (HCoV-HKU1) | NC_006577.2 | β-CoV | |
| 3 | Human coronavirus NL63 (HCoV-NL63) | NC_005831.2 | α-CoV | |
| 4 | Human coronavirus 229E (HCoV-229E) | NC_002645.1 | α-CoV | |

| 5 | Middle East respiratory syndrome-related coronavirus (MERS-CoV) | NC_019843.3 | β-CoV | Caused outbreak in South Korea in 2015 and Saudi Arabia in 2018 |
| 6 | Severe acute respiratory syndrome coronavirus (SARS-CoV) | NC_004718.3 | β-CoV | Caused 2002–2004 SARS outbreak |
| 7 | Severe acute respiratory syndrome coronavirus 2 (SARS-CoV-2) | NC_045512.2 | β-CoV | Cause of ongoing pandemic (2019-) |
| 8 | Bat coronavirus RaTG13 (Bat-CoV) | MN996532.2 | β-CoV | Closest Genome Sequence to SARS-COV-2[6, 34] |

Initially, a phylogenetic tree analysis has been carried out on the downloaded full-length genome sequences. The tree was constructed using the MEGA X software. This is based on maximum likelihood estimation (MLE), Clustal Omega as sequence alignment method with 1000 bootstrap replicates. The Chaos Game Representation map was constructed using the 'kaos' package [35] in R software environment [36] along with the 'seqinr' package [37] for biological sequence analysis. The CGR process results in two series along the *X* and *Y* directions. The signals along the *X* and *Y* directions are then characterized by Fisher-Shannon Information measures. FIM, SEP and FSC are calculated for both *X* and *Y* coordinates. FIM and SEP are calculated for the CGR-*Walk* sequence as well.

## 4. Result

The phylogenetic tree of the eight coronaviruses considered in the present study namely HCoV-OC43, HCoV-HKU1, HCoV-229E, HCoV-NL63, SARS-CoV, MERS-CoV, SARS-CoV-2 and RaTG13 is shown in Fig. 1.

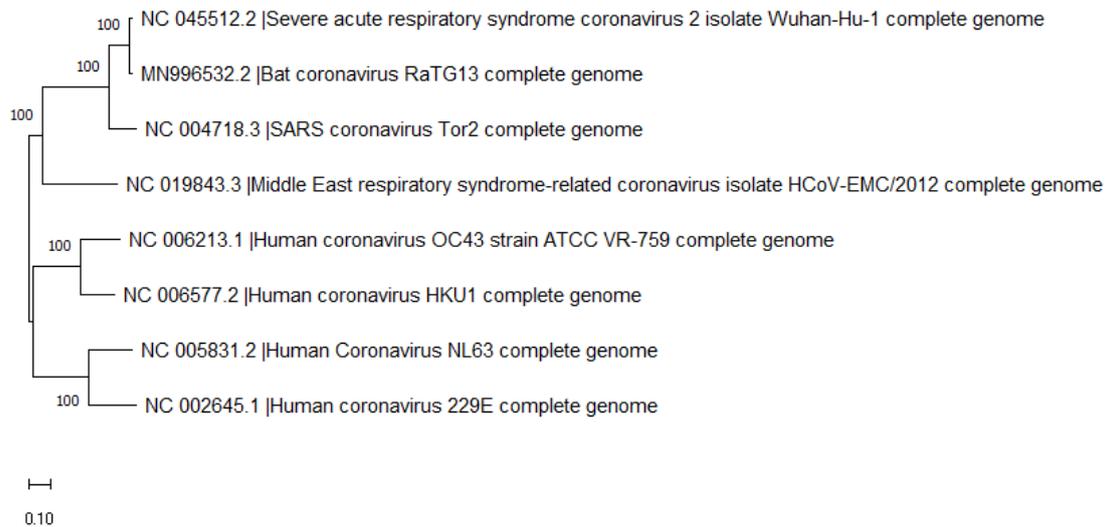

**Fig.1:** Phylogenetic Tree Analysis

From the Phylogenetic Tree in Fig.1 it is evident that SARS-CoV-2 is most similar to RaTG13, followed by SARS-CoV.

The CGR maps of the eight coronaviruses are shown in Fig. 2. Although fractal patterns in the whole genome sequence are discernible in these figures and they look very similar, it is difficult to compare the genomic sequences from visual inspection alone. Therefore, it is important to measure their complexity numerically for a comparative analysis. Thus, as mentioned above the Shannon Power Entropy (SEP), Fisher Information Measure (FIM) and their product Fisher-Shannon Complexity (FSC) are adopted for numerical characterization. These values can be then used for a comprehensive similarity/dissimilarity analysis of the whole genome sequences.

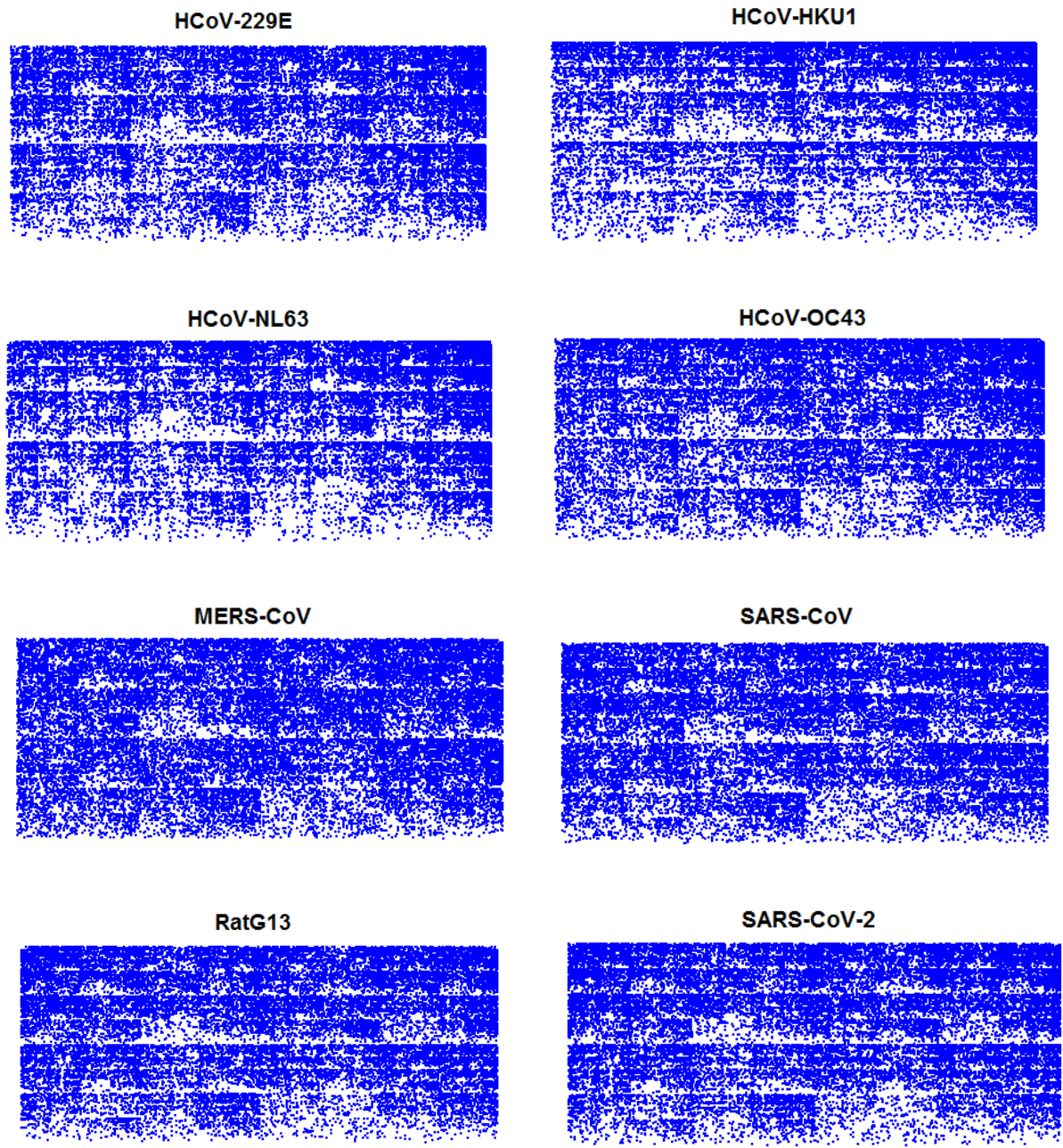

Fig. 2: CGR maps of the coronaviruses

The Fisher-Shannon Information Plane for the *X* and *Y* coordinates of the above mentioned viruses are shown in Fig. 3 and 4 respectively. From the FSIP along *X* coordinates (Fig. 3) it can be seen that HCoV-OC43, RaTG13 and SARS-CoV-2 form a close cluster. Thus, it may be inferred that these viruses are very similar for the CGR-*X* coordinates. The horizontal stripes i.e. the *X*-coordinates give the dinucleotide similarities of AC, CA, GT and TG.

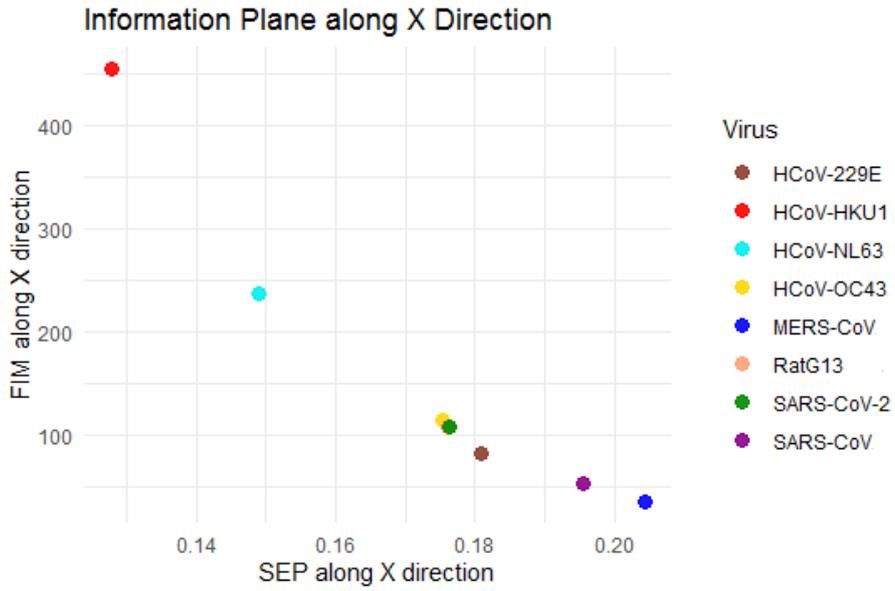

**Fig. 3:** Fisher-Shannon Information Plane along *X* direction (*FSIP-X*)

Also, from the *FSIP-Y* map, it can be seen that MERS-CoV, RaTG13, SARS-CoV-2 and SARS-CoV form a close cluster. Thus, these viruses are very similar for the *CGR-Y* coordinates.

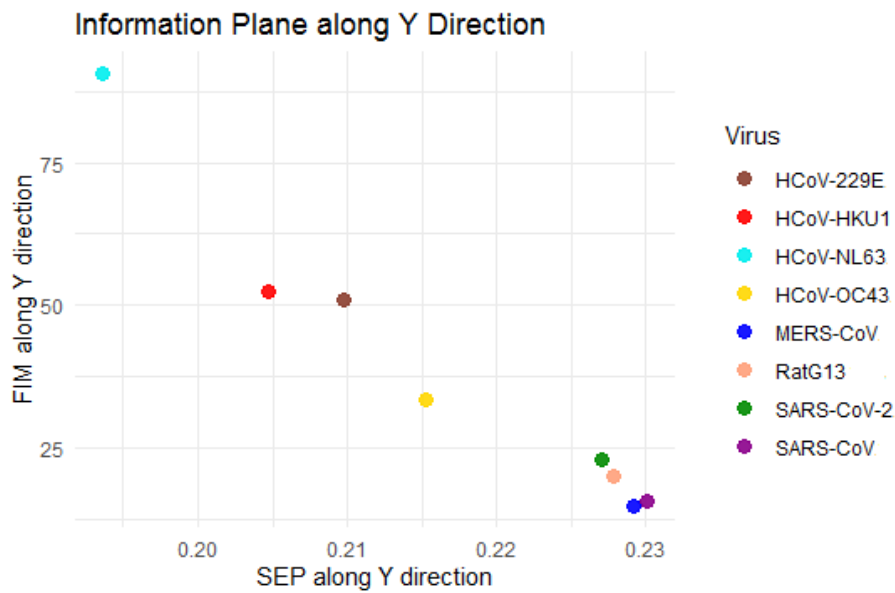

**Fig. 4:** Fisher-Shannon Information Plane along *Y* direction (*FSIP-Y*)

Accordingly, Fisher-Shannon Complexity (FSC) for both *X* and *Y* coordinates of the coronaviruses are plotted in Fig. 5. Here also, RaTG13 and SARS-CoV are very nearby which indicates that they are genetically very similar. Also, the most distant coronavirus from SARS-CoV-2 is found to be HCoV-NL63, which can be observed in the phylogenetic tree as well. Also, it worth mentioning that FSC values are not equal on unity, which shows the non-Gaussian nature of the signal.

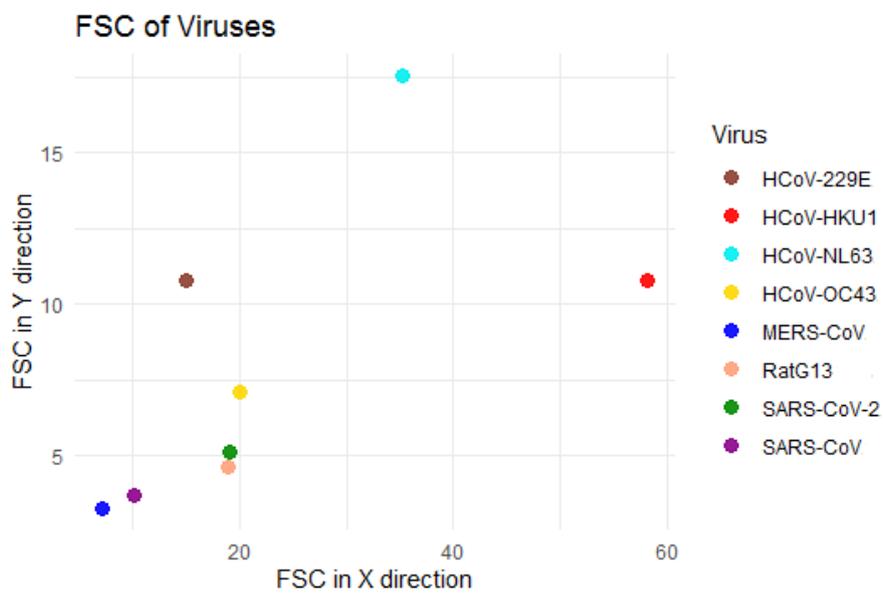

**Fig. 5:** Fisher-Shannon Complexity (FSC) of coronaviruses

In the previous analysis (i.e, Figs. 3, 4 and 5) the quantities like SEP, FIM and FSC are calculated from probability density functions (pdfs) of the *CGR-X* and *CGR-Y* values by treating them as independent random variables. Possibly, more robust quantification of the CGR may be accomplished by assuming a bivariate pdf consisting of random variables along both *CGR-X* and *CGR-Y* and then estimating SEP and FIM from that bivariate distribution. Finally, we also obtain the FSIP of the CGR-walk, which is defined earlier as a summation of the *X* and *Y* coordinates. This results in one dimensional time series. The FSIP is shown in Fig. 6 and it can be again seen that SARS-CoV-2, SARS-Cov and RaTG13 are very nearby which indicates that they are genetically very similar. Thus, from the Figs. 3-6 it can be concluded that the virus SARS-CoV-2 bears most close resemblance to the batcoronavirus RaTG13 followed by SARS-CoV. This confirms the earlier reported results that SARS-Cov-2 genome bears 79.6% sequence similarity to SARS-CoV and 96% similarity to the bat coronavirus [34].

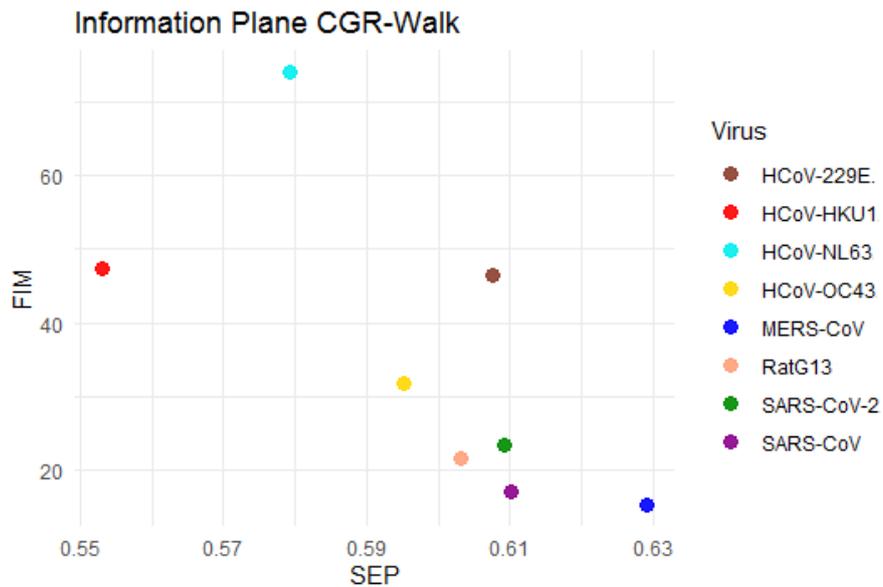

**Fig. 6:** Fisher-Shannon Information Plane for CGR-*Walk*

**5. Conclusion**

In this paper, comparative genome analysis of eight coronaviruses namely HCoV-OC43, HCoV-HKU1, HCoV-229E, HCoV-NL63, SARS-CoV, MERS-CoV, SARS-CoV-2 and RaTG13 is carried out using Chaos Game Representation and Fisher-Shannon Complexity (CGR-FSC) method. The genomic divergence of viruses can be observed in the Fisher Shannon Information Plane (FSIP). The results indicate that the coronavirus SARS-CoV-2 causing COVID-19 in humans is most similar to the horseshoe batcoronavirus RaTG13. The CGR-FSC method is shown to be effective for carrying out a similarity/dissimilarity analysis of DNA sequences.

**Declaration of competing interest**

The authors declare no conflict of interest.

**Funding**

This research did not receive any specific grant from funding agencies in the public, commercial, or not-for-profit sectors.